\documentclass{emulateapj}

\usepackage{epsfig}
\usepackage{epstopdf}

\newcommand{\etal}{{\it et\thinspace al.}\ }

\newcommand{\simlt}{\ {\raise-.5ex\hbox{$\buildrel<\over\sim$}}\ }

\submitted{Submitted 23 December 2008; Accepted 20 February 2009 -- To appear in 10 April 2009 ApJ Letters}

\lefthead{Salzer \etal}
\righthead{High z Low Z Galaxies}

\begin{document}

\title{A Population of Metal-Poor Galaxies with $\sim$L* Luminosities at Intermediate Redshifts}

\author{John J. Salzer\altaffilmark{1,2}, Anna L. Williams\altaffilmark{2}, \& Caryl Gronwall\altaffilmark{3}}

\altaffiltext{1}{Department of Astronomy, Indiana University, Bloomington, IN\ \ 47405; slaz@astro.indiana.edu}
\altaffiltext{2}{Astronomy Department, Wesleyan University, Middletown, CT\ \ 06459; alwilliams@wesleyan.edu}
\altaffiltext{3}{Department of Astronomy \& Astrophysics, Pennsylvania State University, University Park, PA\ \ 16802; caryl@astro.psu.edu}


\begin{abstract}
We present new spectroscopy and metallicity estimates for a sample of 15
star-forming galaxies with redshifts in the range 0.29 -- 0.42.  These objects
were selected in the KPNO International Spectroscopic Survey (KISS) via their
strong emission lines seen in red objective-prism spectra.  Originally thought to
be intermediate-redshift Seyfert 2 galaxies, our new spectroscopy in the far red 
has revealed these objects to be metal-poor star-forming galaxies.  These galaxies
follow a luminosity-metallicity (L-Z) relation that parallels the one defined by low-redshift
galaxies, but is offset by a factor of more than ten to lower abundances.  The amount of 
chemical and/or luminosity evolution required to place these galaxies on the
local L-Z relation is extreme, suggesting that these galaxies are in a very
special stage of their evolution.  They may be late-forming massive systems,
which would challenge the current paradigm of galaxy formation.  Alternatively,  
they may represent intense starbursts in dwarf-dwarf mergers or a major infall
episode of pristine gas into a pre-existing galaxy.  In any case, these objects 
represent an extreme stage of galaxy evolution taking place at relatively low redshift.  
\end{abstract}



\section{Introduction}

It has recently become possible to obtain metallicity estimates for galaxies
with significant lookback times.  While nearly all precision studies of galaxian 
abundances carried out before 2000 focused on nearby objects 
(z $<$ 0.1), recent emphasis on deep, high-redshift surveys has opened the 
way to studying chemical evolution out to z = 1 and beyond (Lilly et al. 2003;
Kobulnicky et al 2003; Liang et al. 2004; Maier et al, 2004, 2005, 2006; 
Shapley et al. 2005; Lamareille et al. 2006; Erb et al. 2006, Kakazu, Cowie \& 
Hu 2007; Liu et al. 2008; Maiolino et al. 2008; Rodrigues et al. 2008).  It is now 
possible to utilize measurements of the chemical evolution of galaxies (e.g., via 
the redshift dependence of the mass-metallicity or luminosity-metallicity (L-Z) 
relation) as an important tool for understanding galaxian evolution in general.  
Any model/scenario for the formation and evolution of galaxies must account 
for the L-Z relation and its change as a function of redshift.

Advances in our knowledge of galaxian metal abundances have also 
occurred among low-redshift samples.  The seminal paper by Tremonti et al.
(2004) using the Sloan Digital Sky Survey (SDSS; York et al. 2000; 
Adelman-McCarthy et al. 2008) provided L-Z and mass-metallicity relations 
for $\sim$53,000 SDSS galaxies.  On a smaller scale, 
Salzer et al. (2005b) utilized the spectra for 765 star-forming galaxies from
the KPNO International Spectroscopic Survey (KISS) to create optical and
NIR L-Z relations.   The availability of reliable local L-Z relations, based on
large, homogeneous samples, has been essential for establishing a baseline
for comparing with high redshift samples.  Much work continues with these
low-z samples in order to better quantify the shape of the L-Z relation, as
well as to improve the calibration of the metallicity scale at the high-abundance
end.

The amount of metallicity evolution occurring between z of 0.5 - 1.0 and
today inferred from the studies mentioned above is small.  This  was
generally expected, and is consistent
with the picture that the bulk of galaxy assembly took place prior to z = 1.  
For example, Lilly et al. 2003 find that only a $\sim$20\% change in O/H
relative to galaxies at z = 0 is required to explain the observed metallicities 
in their sample at a mean redshift of 0.75.   Other studies (e.g., Maier et al.
2004; Liang et al. 2004, Lamareille et al. 2006) infer a larger amount of
metallicity evolution, but in nearly all cases no more than a factor of two for
galaxies between z $\sim$ 1 and today.  However, our recent
observations of KISS galaxies at intermediate redshifts (z = 0.29 - 0.42)
have revealed that the overall picture may not be as simple as previously
believed.  The current paper presents new results that indicate the presence
of a population of very metal-poor yet luminous galaxies at modest lookback
times of 3-4 Gyr.



\section{Higher-z KISS Galaxies}

The KISS galaxy sample is derived from a wide-field Schmidt
survey that selects emission-line objects via the presence of H$\alpha$
emission in their objective-prism spectra (Salzer et al. 2000, 2001; Gronwall
et al. 2004b).   The survey method included using a filter that restricted the 
wavelength coverage of the slitless spectra to 6400--7200 \AA.
KISS detects objects via their  H$\alpha$ emission out to z = 0.095.  However,
follow-up spectroscopy of KISS objects (Wegner et al. 2003; Gronwall et al. 2004a; 
Jangren et al. 2005; Salzer et al. 2005a) revealed that $\sim$2\% of the
overall sample were not detected by H$\alpha$ emission but instead were 
higher-redshift objects with [O III] $\lambda$5007 shifted into the bandpass
of the survey filter.  These objects have z between 0.29 and 0.42.
Virtually all of the 38 [O III]-selected objects exhibited high-excitation spectra
([O III]/H$\beta$ $>$ 3), and most had M$_B$ between $-$19.5 and $-$22.5.
The combination of high-excitation spectra and high luminosities caused us
to classify them initially as Seyfert 2 galaxies, based solely on their blue
rest frame spectra (e.g., see Fig 4 of Gronwall et al. 2004a). 

The activity type of an emission-line galaxy
(e.g., AGN vs. star forming) cannot be unambiguously determined using a
single flux ratio such as [O III]/H$\beta$.  The redshifts of the [O III]-selected
KISS galaxies meant that the key diagnostic line ratio [N II]/H$\alpha$ was
nearly always beyond the red end of our quick-look follow-up spectra.  In order
to resolve the ambiguity regarding their type classifications, we undertook a 
program of spectroscopy in the far red (6300--9500 \AA) that allowed us to 
observe the important lines around H$\alpha$.  All observations for this program 
were obtained using the 9.2-m Hobby-Eberly Telescope\footnote{Based on 
observations obtained with the Hobby-Eberly Telescope, which is a joint project 
of the University of Texas at Austin, the Pennsylvania State University, Stanford
University, Ludwig-Maximilians-Universit\"at M\"unchen, and Georg-August-Universit\"at
G\"ottingen.} 
(HET).
Figure 1 shows spectra of three KISS galaxies observed with the HET.  The details 
of the observations will be presented in Williams, Salzer \& Gronwall (2009, in preparation).

\begin{figure*}[h]
\epsfxsize=4.0in
\epsscale{0.6}
\plotone{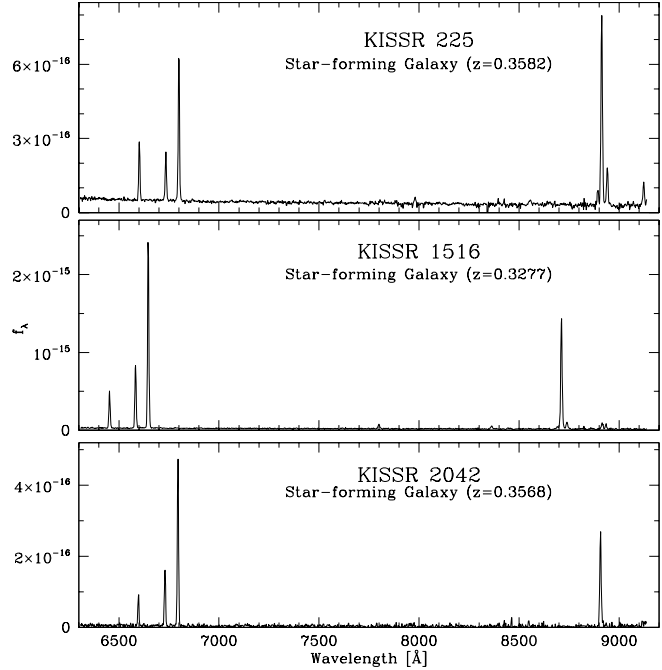}
\vskip -0.01in
\figcaption{Example spectra of [O III]-detected star-forming KISS galaxies as
described in the text.  All spectra were obtained with the HET.  The lower two spectra
are reminiscent of BCDs in the local universe.
\label{fig:spec}}
\end{figure*}

The results of the red spectroscopy were surprising.  As expected, many of the 
[O III]-selected KISS galaxies (23 of the 38, or 61\%) were {\it bona fide} AGNs - mostly 
Seyfert 2s.  However, 15 (39\%) exhibited low [N II]/H$\alpha$ ratios, indicative of gas 
that is photoionized by hot stars.  The three spectra illustrated in Figure 1 are all
examples of the latter.  Figure 2 shows the standard line ratio diagnostic
diagram for the [O III]-selected objects.  The Seyfert 2's are located in the expected
region of the diagram (Baldwin, Phillips \& Terlevich 1981; Veilleux \& Osterbrock
1987).  The star-forming subset of the high z KISS galaxies are located in the 
upper left portion of the diagram, in the area occupied by metal-poor star-forming
dwarf galaxies (a.k.a. blue compact dwarfs (BCDs)).  In other words, the star-forming
galaxies detected via their strong [O III] emission appear to have spectra similar
to low-metallicity dwarf galaxies, despite having luminosities comparable to L*.

\begin{figure*}[htp]
\vskip -0.9in
\epsfxsize=5.0in
\epsscale{0.6}
\plotone{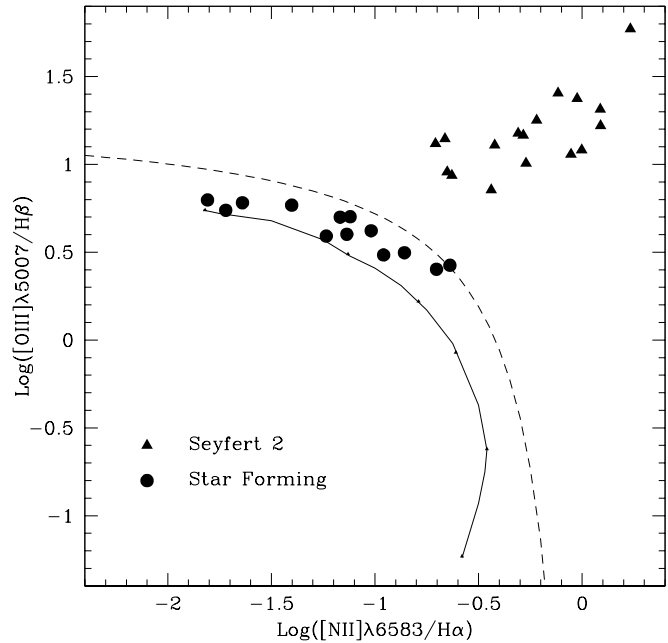}
\vskip -0.01in
\figcaption{Line diagnostic diagram plotting the [O III]/H$\beta$ line ratio against
the [N II]/H$\alpha$ ratio for the higher-z KISS galaxy sample.  The [O III]-detected
star-forming galaxies are plotted as filled circles, while the higher-z Seyfert 2's are
represented by filled triangles.  Only objects with reliable line ratios are included in
the figure.  The dashed line, from Kauffmann et al. (2003), is an empirical demarcation 
betwen AGN and star-forming galaxies.  Our independent classifications are consistent
with this dividing line.  The solid line shows the locus of model HII 
regions with varying metallicity (Dopita \& Evans 1986).  The star-forming galaxies 
are located in the metal-poor part of the diagram, despite having high luminosities.  
\label{fig:diagplot}}
\end{figure*}


\section{Low-Metallicity Star-Forming Galaxies}

We estimate the oxygen abundances of the [O III]-selected star-forming galaxies
using the ``coarse abundance" method described in Melbourne \& Salzer (2002) and
Salzer \etal (2005b).   This method utilizes ratios of strong lines ([O III]$\lambda$5007/H$\beta$, 
[N II]$\lambda$6583/H$\alpha$) present in the spectra to infer the abundance,
based on an empirical calibration between the ratios and metal abundances for
objects of known metallicity.
While more accurate estimates are possible for many of the higher-z KISS galaxies,
we employ the coarse method here to be consistent with the method
used to estimate the abundances of the comparison sample of low-z KISS
galaxies.  As emphasized by Kewley \& Ellison (2008), it is essential to use a
consistent metallicity calibration when comparing different L-Z relations.
 The uncertainty in the O/H values derived in this way is 0.15--0.20 dex 
(Salzer \etal 2005b).  Key properties of the [O III]-selected KISS galaxies, including
their luminosities and abundance estimates, are given in Table 1.  The B-band
absolute magnitudes were derived using the V-band photometry published as
part of KISS (e,g,, Salzer et al. 2001), since V is fairly close to rest frame B for
the redshifts of interest.  All KISS photometry is on the Vega system.  A small 
incremental K-correction  was then applied using 
the observed color and precise redshift of each KISS galaxy.  Note that two of the
galaxies in the sample exhibit spectra with S/N ratios too low to provide a reliable
abundance estimate.  These galaxies are not included in the following discussion.

Figure 3 presents an L-Z diagram for the 13  [O III]-selected star-forming galaxies.
Also plotted are 1363 low-z KISS galaxies ($<$z$>$ = 0.058, z range = 0.0 to 0.095)
with metallicity estimates derived in an
identical way.   The upper solid line is a bivariate linear fit to the low-z galaxies, while
the lower dashed line has the same slope but is offset to provide the best fit to the
higher z objects.  The fit to the [O III]-selected galaxies is offset by $\sim$1.1 dex to 
lower abundances (factor of 13) relative to the low-z sample.  This is an amazing result, 
since it implies that up to a factor of 13 increase in the metallicity of these galaxies is 
required between a redshift of 0.29--0.42 and today (a lookback time of 3-4 Gyr, where 
we assume H$_o$ = 70 km/s/Mpc, $\Omega_M$ = 0.27, and $\Omega_\Lambda$ = 0.73). 
A key point to stress is that none of the galaxy samples referred to above with redshifts 
out to z = 1 and beyond exhibit such a large metallicity offset.  Most would be no more than 
a factor of two below the local L-Z relation (0.3 dex), despite having much larger lookback times.
It is worth noting that the large metallicity offset seen in Figure 3 between the two samples 
is present regardless of which abundance calibration used by Salzer et al. (2005b) is adopted.

As mentioned above, the spectra of the [O III]-selected star-forming galaxies closely
resemble those of low redshift BCDs.  The lower two spectra presented in Figure 1
are particularly good examples of this fact.  It is tempting to interpret their
locations in the L-Z diagram as being due to an extreme luminosity enhancement
due to a starburst in a dwarf galaxy.   If the offset of the 
[O III]-selected galaxies from the mean trend defined by the low-z KISS galaxies is 
assumed to all be due to an increase in their luminosity, it would require an
increase of 3.6 magnitudes (factor of $\sim$28)!   For BCDs in the local universe,
the luminosity enhancement from their current starburst is more typically a factor
of two above the luminosity of the quiescent host galaxy (e.g., Salzer \& Norton 1999;
Lee et al. 2004).   A more plausible
interpretation would be that the high z KISS galaxies are {\it both} metal-poor and
over-luminous relative to local star-forming galaxies.  If we assume a factor
of two luminosity enhancement due to the starburst (0.75 mag), then the implied
offset in metallicity from the local sample is  $\sim$0.9 dex, or a factor of $\sim$8.
This is still a factor of 4 below the L-Z relations of the z $\sim$ 1 galaxies referred
to above.


\section{Discussion}

The discovery of a population of low-abundance galaxies with $\sim$L* 
luminosities at intermediate redshifts raises two fundamental questions.  
First, why haven't these types of objects been discovered previously? 
Second, what is the nature of these enigmatic objects?

\subsection{A New Class of Galaxy?}

Given the large number of studies of metal abundances in galaxies with 
intermediate and high redshift mentioned in the Introduction, it may seem 
odd that systems similar to those described here have not been recognized
previously.  Our interpretation of this enigma centers on the fact that the
[O III]-selected low-metallicity KISS galaxies are likely to be quite rare.  
Over the 136.1 deg$^2$ covered by KISS lists included in this study, only 
15 such objects were detected.  Hence, the probability of finding an analog 
to one of these systems in a deep pencil-beam survey 
(e.g., GEMS (Giavalisco et al. 2004), GOODS (Rix et al. 2004), etc.) 
that tend to cover a small fraction of a square degree is small.  Furthermore,
the typical apparent magnitude of these sources is B $\sim$ 20-22, making 
them fainter than the spectroscopic limits of wide-field surveys like SDSS.   

The selection method of KISS favors the detection of objects 
with strong, high-equivalent-width emission lines.  Even objects with no
detectable continuum (and hence very faint apparent magnitudes) can
be detected in KISS if their emission lines are of sufficient strength.  
Interestingly, the types of objects described here are probably the {\it only}
type of star-forming galaxy that KISS could be sensitive to at these redshifts.
Metal-rich star-forming systems would have [O III] lines that are too weak
for detection, while strong-lined dwarfs would likely be too faint.  Only this
type of $\sim$L* galaxy with low metal abundances and hence strong
[O III] lines could be detected by KISS in this redshift range.

We do not wish to imply that the objects described here are unique.  Rather, 
they have largely been missed by previous studies.  Even so, there are 
examples of similar galaxies in the literature.  For example, the study of Maier
et al. (2004) includes one object in their z $\approx$ 0.4 sample that is well below
the L-Z trend defined by the rest of the sample.  More recently, Kakazu et al. (2007)
used a narrow-band selection process to select strong emission-line galaxies via
the [O III] line at z = 0.63 and 0.83.  Several of their objects fall well below the local
L-Z relation, and may be examples of the type of galaxy we are describing here.

Another sample of galaxies that appear to bear some resemblance to the 
[O III]-selected KISS galaxies are the so called local Lyman Break analogs 
studied recently by Hoopes et al. (2007) and Overzier et al. (2008).  These
objects are selected primarily on their UV properties, and have redshifts of
0.1 to 0.3.  While these objects have been found to be slightly metal poor 
compared to normal galaxies of the same mass (by roughly a factor of 2 for 
galaxies in the luminosity range covered by the KISS sample), they are not
as extreme as the [O III]-selected KISS galaxies.  Additional study will be
needed to determine whether the two samples are drawn from the same
population of galaxies.

A lower limit to the volume density of this class of galaxy is 4.3 $\times$ 10$^{-7}$ 
Mpc$^{-3}$, obtained by dividing our 15 detections by the effective volume of the 
KISS survey in the z = 0.29 - 0.42 range.  This assumes that all of the detected
sources would be visible throughout this volume.  In reality, the actual density
is likely to be somewhat higher, but the small sample size makes deriving a
more accurate effective volume dubious.  To compare our density estimate
with that of the overall galaxy population at this redshift, we utilize the luminosity
function derived by Faber et al. (2008) for galaxies in the redshift range 0.2 -- 0.4.  
Integrating over absolute magnitudes of M$_B$ = $-$19.8 to $-$21.8, we obtain an 
overall volume density of 2.5 $\times$ 10$^{-3}$ Mpc$^{-3}$.  Hence, the [O III]-selected 
low-metallicity KISS galaxies appear to make up only a tiny fraction of the galaxy 
population in this redshift range.  

In the Kakazu et al. (2007) study mentioned above, their [O III]-selected samples at  
z = 0.63 and z = 0.83 both have volume densities of order 10$^{-3}$ Mpc$^{-3}$.
When compared to the much smaller value found for the KISS [O III]-detected
sample, this suggests the possibility that strong-lined [O III] emitters were more 
common at earlier epochs.  However, given the different sensitivity limits of the two 
samples, plus the likelihood that the overall properties of the two sets of galaxies are 
vastly different, one should interpret this result with caution.

\subsection{Evolutionary Scenarios}

What exactly do these luminous but metal-poor galaxies represent?   One
possibility is that they are ``straggler" galaxies, essentially the last
group of massive objects to collapse and form stars.  In this case, their low 
abundances are due to their relatively young ages.  While the current paradigm 
of galaxy formation calls for most massive systems to begin their collapse and
subsequent star formation at a much earlier epoch, we cannot
rule out the possibility of late-forming galaxies.  In this scenario, the progenitor
gas clouds initially would have resisted gravitational collapse, 
perhaps due to a low overall gas density.  Subsequent evolution of the system
might have been inhibited until a specific event occurred to raise the density
sufficiently for collapse to occur.  Potential triggers could include cloud-cloud
collisions or tidal interactions with a passing galaxy.  Such a scenario would
likely demand that the progenitor system resides in a fairly low-density
region.  Future studies of the local environments of the [O III]-selected KISS
galaxies would likely yield interesting results.  Since metallicity enrichment
can occur fairly quickly in massive galaxies, there would be sufficient time
in the intervening 3-4 Gyr for the abundances to increase, such that these
galaxies would no longer stand out as being metal poor in the local universe.

Alternatively, these objects may represent extreme starbursts in otherwise
fairly normal dwarf galaxies.  In this case, the low metallicities are simply
the norm for the class.  Their locations in the L-Z diagram would be due to 
a large brightening due to a massive starburst.   We showed in $\S$3 that
the luminosity enhancement needed to explain these objects is a factor of 
$\sim$28.   This would only be possible in the case of an unusually severe 
dwarf-dwarf merger.  However, this amount of luminosity enhancement
seems extreme!  In the local universe the typical brightening due to a
starburst in a BCD (the systems with the highest relative starburst strengths)
is only a factor of 2 - 3 in most cases (Salzer \& Norton 1999).  

Another way to explain these systems would be the accretion of large amounts
of pristine gas by L* galaxies that start out with fairly normal metallicities.  However, 
this scenario would require that the mass of accreted gas is roughly a factor of 
ten larger than the mass of the original ISM in order to dilute the gas phase abundances 
to the observed levels.  Again, the likelihood of this happening seems low, but it
can't be ruled out.

Regardless of the cause, these are extremely interesting systems.
The fact that none are observed in the local universe indicates that rapid
evolution of the systems is required.  Presumably, we are witnessing a 
short-lived episode in the life of these objects, one in which they are
going through an extreme phase in their evolution.  Whether or not this 
is a normal process that all galaxies go through at some point remains to
be determined.  There is a distinct possibility that these objects are going
through an evolutionary stage that is common to most galaxies, albeit at an
earlier epoch.  If true, the study of these galaxies will allow us to probe 
a critical phase in galaxy evolution using systems that are much closer than
their high-redshift counterparts.


\acknowledgements
We gratefully acknowledge financial support for the KISS project from an NSF 
Presidential Faculty Award to JJS (NSF-AST-9553020), as well as continued 
support for our ongoing follow-up spectroscopy campaign (NSF-AST-0071114 and
NSF-AST-0307766).
We thank the many KISS team members who have participated in the spectroscopic 
follow-up observations during the past several years, particularly Gary Wegner, 
Drew Phillips, Jessica Werk, Laura Chomiuk, Kerrie McKinstry, Robin Ciardullo, 
Jeffrey Van Duyne and Vicki Sarajedini.  Enlightening discussions with Rose Finn 
and George Helou are gratefully acknowledged.  We wish to thank the support 
staff of the Hobby-Eberly Telescope for their excellent work in obtaining the 
spectroscopic observations that made this work possible.  Finally, we are highly
appreciative of the many excellent suggestions made by the referee.



%
%


\begin{figure*}[htp]
\epsfxsize=5.0in
\epsscale{0.6}
\plotone{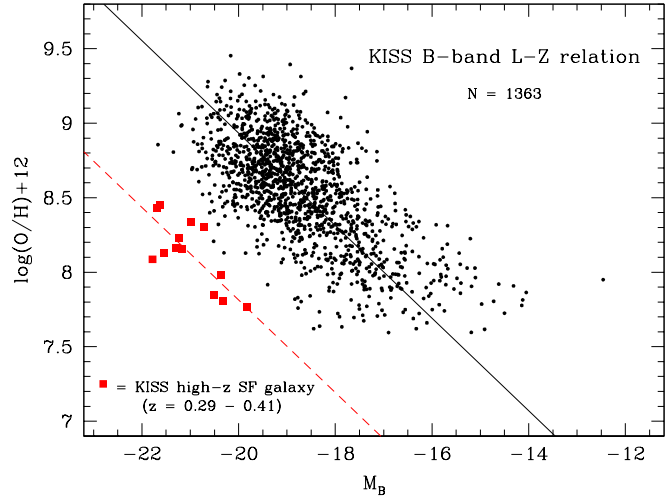}
\vskip -0.01in
\figcaption{Luminosity-metallicity relation for 1300+ low-z KISS galaxies (z $<$ 0.095; small dots) 
and the 13 [O III]-detected star-forming KISS galaxies (red squares).  The solid line is a linear fit 
to the low-z galaxies, while the lower dashed line has the same slope but fits the higher-z galaxies 
with an offset of -1.1 dex.  
\label{fig:lzplot}}
\end{figure*}


\begin{deluxetable}{rcccc}
\tablecaption{[O III]-selected KISS Star-forming Galaxies \label{tab:data}}
\tablehead{
\colhead{KISSR\tablenotemark{a}} & \colhead{z} & \colhead{B} & \colhead{M$_B$} & \colhead{log(O/H)+12} \\
\colhead{(1)}& \colhead{(2)}& \colhead{(3)}& \colhead{(4)}& \colhead{(5)} 
}
\startdata
 169 & 0.4029 & 20.82 & -21.16 & \nodata \\
 225 & 0.3582 & 19.85 & -21.69 & 8.43 \\
 560 & 0.3579 & 21.54 & -20.51 & 7.85 \\
 847 & 0.3557 & 22.00 & -19.83 & 7.76 \\
 980 & 0.3434 & 22.05 & -20.82 & \nodata \\
1038 & 0.4100 & 20.69 & -21.30 & 8.16 \\
1290 & 0.3050 & 20.19 & -21.17 & 8.15 \\
1508 & 0.2940 & 19.99 & -21.23 & 8.23 \\
1516 & 0.3277 & 19.59 & -21.53 & 8.13 \\
1759 & 0.4050 & 21.83 & -20.36 & 7.98 \\
1791 & 0.3592 & 20.49 & -20.72 & 8.30 \\
1825 & 0.3311 & 19.64 & -21.62 & 8.45 \\
1953 & 0.3688 & 20.04 & -21.78 & 8.09 \\
2005 & 0.3081 & 20.10 & -20.99 & 8.33 \\
2042 & 0.3568 & 21.08 & -20.33 & 7.81 \\
 \enddata
\tablenotetext{a}{KISSR stands for KISS Red, the H$\alpha$-selected portion of KISS.  All objects
included in this study are found in the 1st and 2nd H$\alpha$-selected KISS catalogs (Salzer et al. 
2001; Gronwall et al. 2004b)}
\end{deluxetable}


\end{document}